# What do adoption patterns of solar panels observed so far tell about governments' incentive? insight from diffusion models

Anita Bunea[1], Pompeo Della Posta[2], Mariangela Guidolin[3], Piero Manfredi[2]

[1]Institute of Management, Sant'Anna School of Advanced Studies, Pisa, ITALY.

[2]Department of Economics and Management, University of Pisa, ITALY.

[3] Department of Statistical Sciences, University of Padua, ITALY.

**Abstract**

**Background**. The fast worldwide spread of renewable energies is one of the key actions in the international response to the threats of global warming and climate change. However, the diffusion of solar photovoltaic panels (SPP), continuously shows stalling epochs despite wide support from public incentive.

**Objectives.** To improve the general understanding of the main communication determinants of diffusion of SPP worldwide, with special focus on the role played so far by public incentives, and of the resulting future perspectives.

**Methods and data**. By upgrading previous research relying upon data up to 2006 i.e., before the main public interventions were undertaken in most countries, we applied the generalized Bass model to an extended dataset on SPP adoptions (26 countries between 1992-2016) for characterizing the temporal profile of the major domestic shocks in SPP markets, mostly occurred after 2007, focusing on the role of public interventions in influencing scale and shape of SPP adoption curves.

**Results.** (i) The SPP market started everywhere without the assistance of continued public media support, so that its initial lifecycle was sustained by word-of-mouth communication only, (ii) the pace of word-of-mouth was however generally low, suggesting that, overall, in SPP markets communication forces acting on are not effective compared to the time scales which are required to respond to the global threats, (iii) almost everywhere the largest part of market growth resulted from massive positive shocks occurred between 2007 and 2016, possibly following incentive measures in the various countries, iv) the estimates of parameter describing the temporal pattern of the shocks revealed a lack of temporal persistence of the effects of incentive and a sharp trade-off between intensity and persistence of the actions.

**Concluding remarks.** The SPP market appears as a frail and complicate one, inadequately supported by media communication. Public incentives were necessary for the market take-off but they were often badly designed, resulting in late, short-term, responses to external stimuli (such as e.g., the cogent pressure of international deadlines) in the absence of a well-established long-term plan. They are therefore failing to give a sustained momentum to the market. The limited temporal persistency of public actions, causing the market to be dominated by incentive-forced waves followed by negligible post-incentive adoptions suggest the emergence, at least in some countries, of an "addiction to incentive" phenomenon, and therefore a deleterious role of expectation.

## 1. Introduction

The fast worldwide transition towards renewable energies (RE), from solar photovoltaic panels (SPP), to wind energy and biomasses, is one of the critical factors in the fight against the scaring impact of global warming and climate change (Bonn, 2017; IPCC, 2014; Kyoto Protocol, 1998; Paris Agreement, 2015, Armaroli and Balzani, 2010; ITRPV, 2018). Nonetheless, the pathway towards the generalized use of RE is still sown of obstacles which go well beyond the attempts to debunking global climate change from adverse science and policy (McCright and Dunlap, 2011, 2003; Oreskes, 2018). As regards SPP, which is the only RE technology currently available to households, these obstacles are intrinsic to SPP adoption as a long-term investment which is still perceived as unsustainable in view of its high installation costs (Masini and Frankl, 2003; Palmer et al., 2015; Yamaguchi et al., 2013; Zhang et al., 2011), its long payback period (Dharshing, 2017; Islam, 2014; Robinson et al., 2013), the information costs associated with management and maintenance (Vasseur and Kemp, 2015), constraint on financial resources (Palmer et al., 2015; Robinson et al., 2013), and finally, the uncertainties about the future energy policies that governments would be following (Reddy and Painuly, 2004; Vasseur and Kemp, 2011) and about future technological developments (Karneyeva and Wüstenhagen, 2017; Ruby, 2014)**.**

Public incentives therefore represent the primary instrument to stimulate the domestic demand by moderating these costs and helping the SPP market competing successfully against the «dirty» alternatives, which are less costly but not environmentally friendly (Avril et al., 2012; Chowdhury et al., 2014; Ratinen and Lund, 2015; Zhang et al., 2011). Additionally, public intervention can play a key role on the supply side. For instance, an infant SPP industry willing to enter a market which is already dominated by an incumbent foreign producer who is enjoying lower average costs thanks to the large market share it serves, would have a hard time being successful, given the high unitary cost due to the limited scale of its production – rather than by its inefficiency. Public incentives might help removing this obstacle. The supply side can also be favoured by improving the quality of the technology for given costs, which, if allowing to charge higher selling prices, would increase profitability. Other roles of public policies that have not been stressed enough in the literature are the need to provide a viable source of energy for countries with a low endowment of fossil fuels, especially when their market prices increase (Lee and Huh, 2017), and to acquire a technology leadership, as it is appears in all evidence for China which has become the major worldwide producer of solar modules since 2007 (IEA, 2016, p. 53).

The previous reasons have led many countries to introduce incentive measures supporting the SPP market, most of which are customer oriented (IEA, 2016; IEA International Energy Agency, 2018; ITRPV, 2018). Nonetheless, the scale and pace of diffusion of SPP are still too low (Karneyeva and Wüstenhagen, 2017) - despite the almost continuing decrease of prices (ITRPV, 2018)- compared to what would be necessary to respond effectively to current societal challenges. In turn, this raises doubts as to the true obstacles discouraging the growth of the domestic demand, but also as to the role played by incentives and by the way they are assigned, that might have made the SPP market totally dependent on them and therefore being unable to develop any autonomous self-sustained diffusion pathway.

The main objective of this paper is to improve our understanding of the key determinants of SPP worldwide diffusion, and to offer perspectives on the future development of the market with special focus on the role of public policy. Accordingly, we used diffusion models to analyse an extended dataset on SPP adoptions including 26 countries covering about 98% of the worldwide SPP diffusion during 1992-2016. In the management sciences, the use of diffusion models to study the temporal shape of adoption patterns of new durables and technologies was fuelled by the celebrated Bass model (BM, Bass, 1969). The success of the Bass model was due to two main reasons. The first one lies in its clear causal mechanism, identified in the social communication forces, namely media communication on the one hand, and imitation, or *word-of-mouth*, along spontaneous communication between agents ( Mahajan et al., 2000, 1990), on the other hand. The second one lies in its ability to parsimoniously describe observed adoption trajectories, by using only social communication parameters (Bass et al., 1994; Mahajan et al., 2000, 1990). The basic BM was later extended into the so called generalized Bass model (GBM, Bass et al., 1994) to include the effects of marketing decision variables such as prize and advertising, represented through exogenous time-dependent perturbations of diffusion parameters. However, as emphasized by Guseo and co-workers (Guidolin and Guseo, 2016, 2014; Guidolin and Mortarino, 2010; Guseo et al., 2007), the main strength of the GBM lies in its ability to incorporate in a manageable form external shocks - including, among other, the state interventions- forcing diffusion trajectories out of their natural pathway induced by communication forces. More in detail, in this work we apply the GBM to characterize (i) the mutual role played by the communication drivers, namely the media and word of mouth, vs that of public incentives in influencing scale and shape of SPP adoption curves, (ii) the temporal pattern of the major shocks in SPP markets, which mostly occurred after 2007.

This work draws inspiration from, and upgrades, previous work by Guidolin and Mortarino (2010) who first used the GBM to highlight the effects of public incentive in the eleven countries which mostly contributed to SPP worldwide adoptions till 2006. There are two main reasons why we found appropriate to upgrade that paper. The first one is that it has been only after 2007 that the world SPP market definitely took-off, adding more than 95% of the total SPP capacity so far installed worldwide (IEA 2018). This meant an unprecedented growth even in those countries, as Germany and Japan, which had already experienced a significant adoption history. The second one lies in the nature of SPP markets. Indeed, a conclusion in Guidolin and Mortarino (2010) was that some "early" SPP markets, such as Japan, the Netherlands and the UK, had already entered their maturity phase. This conclusion, though correct based on the adopted model, was soon denied by reality, which instead showed since 2007 a dramatic growth in SPP adoptions in all considered countries, possibly resulting from major policy efforts aimed to sustaining the domestic SPP demand initiated synchronously since 2007 in most countries and. We therefore thought that, far from representing a forecasting failure of the Bass model, this was instead evidence of the complexity prevailing in the SPP market, which deserved an updated assessment of the main determinants, and barriers, to SPP adoptions, as well as of the impact of incentive policies, with a special focus on their ability to persist over time and to bring final momentum to the market.

The rest of the paper is organized as follows. Section two presents the methodology and the data. The main general results of the application of the GBM are reported in section three. Concluding remarks follow. Further information, including full details on the results and a discussion of individual countries results in the light of the underlying national energy framework and adopted policies, are reported in the online appendix.

## 2. Methods and data.

*2.1. The Bass model for innovation diffusion*

The original Bass model (Bass, 1969) describes an irreversible diffusion process where an item (a durable good, a new technology, a new idea, etc) spreads in a fixed population of potential adopters having size *m*, as a consequence of the action of the main communications channels. These are distinguished into (i) the "internal" channel, or *word-of-mouth,* following the spontaneous communication between individuals as a consequence of their daily social encounters (be they real or virtual), and (ii) the "external" channel, following the continued individual's intrinsic propensity to adopt from the publicly available information by the media and the public system. The model is described by the following differential equation in the absolute cumulative number of adopters at time t, Y(t):

$$Y'(t) = \left(\alpha + \frac{q}{m}Y(t)\right)(m - Y(t)) \quad (1)$$

where $\alpha>0$ and $q>0$ respectively represent the *innovation coefficient,* tuning the intensity of the external channel, and the *imitation coefficient* tuning the intensity of internal communication. Note that *m* represents the saturation level of the market and is therefore also referred to as the *market potential*. The prime derivative of the cumulative function S(t)=Y'(t) represents the instantaneous adoption rate i.e., the absolute incidence of new adoptions per unit of time. Letting F=Y/m to denote the relative cumulative adoption curve, the corresponding hazard rate is given by:

$$\lambda(t) = \frac{F'(t)}{1-F(t)} = \alpha + qF(t) \quad (2)$$

Equation (1) has the "natural" initial condition Y(0)=0, corresponding to the situation where no adopters exist at the moment where public communication starts. The resulting solution of (1) is:

$$Y(t) = m \frac{1-e^{-(\alpha+q)t}}{1+\frac{q}{\alpha}e^{-(\alpha+q)t}} \quad (3)$$

Equation (3) depicts a monotonically increasing S-shaped curve for $q > \alpha$ and a concave one in the opposite case. In both cases the market relative growth rate $r(t) = Y'(t)/Y(t)$ is monotonically decreasing. In particular, for $\alpha = 0$ the Bass model simplifies into a *pure imitation*, or *internal*, model. In this case $r(t)$ is essentially constant in the initial stages of the market, mirroring an underlying exponential growth of cumulative adoptions. For $q = 0$, the Bass model simplifies into purely external model driven by media communication only.

*2.3 The "generalized" Bass and internal models*

The GBM extends the Bass model by including a general multiplicative, time-dependent, factor *h(t)* in the hazard rate, yielding to the equation:

$$Y'(t) = h(t)\left(\alpha + \frac{q}{m}Y(t)\right)(m - Y(t)) \quad (4)$$

The general solution of (4) for an arbitrary initial condition $Y(0)=Y_0$ is given by

$$Y(t) = mF(t) = m \frac{1 - e^{-(\alpha+q)\int_0^t h(s)ds} \rho_0}{1 + \frac{q}{\alpha} e^{-(\alpha+q)\int_0^t h(s)ds} \rho_0}, \quad \rho_0 = \frac{\alpha(m - Y_0)}{\alpha m + qY_0} \quad (5)$$

The initial condition can be taken as a further parameter to be estimated. In Guidolin and Mortarino (2010) a different but equivalent representation was used by incorporating a time span (of length *d* to be estimated) between the true initialisation of the process (for Y=0) and the moment when the first positive observed datum on adoptions becomes available. For practical purposes it is convenient to represent function $h(t)$ in the form of additive perturbation i.e., as $h(t) = 1 + g(t)$ ( Guseo et al., 2007), where *g* represents the "shock" function. For *g=0* at all times, the GBM reduces to the basic Bass model, while the case $g(t) \geq 0$ ($g(t) \leq 0$) over a given time interval describes a positive (negative) shock. For negative shocks it is necessary to add the condition that the time average of $g(t)$ must always exceed (-1) to preserve the non-decreasing character of the cumulative function. For $\alpha = 0$, the GBM collapses into a model driven by imitation and its perturbations, that we will term a *generalised internal model* (GIM). This model is of interest here because it will prove to be the appropriate model for SPP data in all countries considered. In particular, the relative rate of growth of the GIM model is given by:

$$r(t) = \frac{Y'(t)}{Y(t)} = h(t)q\left(1 - \frac{Y(t)}{m}\right) \quad (6)$$

i.e., it is proportional to the word-of-mouth coefficient (q) times the shock function times the fraction that has not yet adopted.

*2.4 The shock function and its parametrization*

For empirical analyses function *g* can be specified by appropriate parametric forms depending on some parameters vector $\vartheta$ which can be estimated jointly with ($\alpha$, *q*, *m*). In Guidolin and Mortarino (2010) both the constant shock function over an interval (form 1, F1)

$$g(t) = A \cdot I_{\{a,b\}}(t), \quad \vartheta = (A, a, b) \quad 0 < a < b, \ A \in R \quad (7)$$

and the exponential shock function (form 2, F2)

$$g(t) = A \cdot e^{-c(t-a)} I_{(a,\infty)}(t), \ 0 < a, A \in R, c \in R \quad (8)$$

were considered, where $I_{(a,b)}(t)$ represents the indicator function of interval (a,b). Form F1 mirrors a shock uniformly affecting communication parameters ($\alpha$,q) during a certain interval of time, while F2 describes a shock which initiates abruptly and subsequently decays, or inflates, exponentially over time. Another convenient form is the following one (F3):

$$g(t) = A \cdot (t-a) e^{-c(t-a)} I_{(a,\infty)}(t) \quad a > 0, c > 0, A \in R \quad (9)$$

which represents shocks (both positive and negative) which emerge gradually, rather than abruptly as F2, before being re-absorbed with an exponential-like pattern. This mirrors the realistic fact that an incentive policy will hardly result in a sudden change in adoption parameters, instead it will take time for a number of reasons e.g., for the policy to be communicated to the public and subsequently to "materialise" the intention to adopt following awareness of incentive into the actual adoption decision.

Forms 1,2,3 represent a single shock phenomenon but can be readily extended to consider generalized shocks functions $g_i$ over different time ranges $I_i$ by considering e.g., $g(t) = \sum_i g_i(t) I_i(t)$, where each $g_i(t)$ term represents a single shock (Guseo et al., 2007). As noted in Guidolin and Mortarino (2010) the GBM is valuable to provide diagnostics of external interventions and to summarize their temporal characteristics such as e.g., effectiveness, time persistency, etc.

*2.5 Parameter estimation, criteria for inclusion of shocks and goodness of fit.*

The vector of model unknown parameters β=($\alpha, q, m, \vartheta$) was estimated from available data by nonlinear least squares (NLS) by considering the standard nonlinear regression model (Seber and Wild, 1989):

$$z(t) = Y(t, \beta, Y_0) + \varepsilon(t) \qquad (10)$$

where the observed response *z(t)* is the sum of the deterministic component, here represented by the GBM cumulative curve in equation (5), and the error term ε(t), taken as a standard white noise error (Seber and Wild, 1989). The white noise hypothesis is a simplifying one but we maintained here for sake of simplicity as it was also used in Guidolin and Mortarino, 2010.

In particular, concerning the market potential, as the direct estimation of *m* was feasible only in a few cases, we had to take *m* as a fixed parameter. In the main text we focused on the more relevant case where m is set to the nearest short-term target (available for essentially all countries considered) established for the SPP market by the underlying state Energy Authorities. This short-term scenario will also be referred to as the "minimum" scenario. In the online appendix also results from a scenario with m fixed at a long-term target (when available) are reported.

The inclusion of shock terms in the GBM was stepwise along the following procedure: (i) visual inspection of the data, (ii) preliminary fit of a basic Bass model and examination of regression residuals; (iii) stepwise inclusion of shocks, to maintain parsimony, by accepting the next shock based on the incremental values of the squared multiple partial correlation coefficient (SMPCC), as in Guidolin and Mortarino (2010). The latter measure is useful to evaluate the improvements in model performance during the stepwise procedure from the "current" model with (i-1) shock terms, to the "next" one with i shocks (Guidolin and Mortarino, 2010; Guseo and Guidolin, 2009) and is defined as follows:

$$\tilde{R}^2_{i-1,i} = \frac{R^2_i - R^2_{i-1}}{1 - R^2_{i-1}} \qquad (11)$$

where $R_i^2$ represents the determination index of model i. Measure (11) captures the relative reduction of residual deviance achieved through the fitting of the next GBM. As a rule, a model is considered to better explain the SPP trend if $\widetilde{R}_{i-1,i}^2$ is larger than an appropriate threshold. Here we set this threshold to 0.5 but results were robust to alternative choices.

*2.6 Data*

Yearly data over the period 1992-2016 on cumulative installed SPP capacity (in MegaWatt) in the 26 countries considered were gathered from published international sources (IEA; IRENA). The installation data cover the eleven countries included in the analyses in Guidolin and Mortarino (2010) (Australia, Austria, Canada, France, Germany, Italy, Japan, the Netherlands, Spain, the UK and the US) and fifteen additional countries (Belgium, China, Denmark, Finland, Israel, South Korea, Malaysia, Mexico, Norway, Portugal, Sweden, Switzerland, Thailand, and Turkey) overall representing 96.5% of the total worldwide capacity installed up to December 2016.

Data on National targets for SPP (for each country considered) drawn from a range of published sources from national energy authorities (see Appendix, Table A. 1) were used to define the short-term target for the market potential m.

**3. Results**

We preliminary remark that, compared to the work by Guidolin and Mortarino (2010), who analysed the time window 1992-2006 when SPP adoption was in most countries in an early stage, our extended data show a number of further "candidate" shocks during 2007-2016. These shocks were characterized by a much larger magnitude than those observed before 2006, not only in absolute terms – which would be expected – but also in relative terms despite the larger scale achieved by the market. Due to the need to include a number of shocks during the period 2007-2016, and therefore facing a rapidly increasing number of parameters to be estimated, we preferred - with a few exceptions (notably Germany and Japan) - not to include further shocks in the first stages of SPP lifecycle i.e, prior to 2007. This was motivated by the fact that almost everywhere the early growth rates of cumulative installations indicated a coarsely constant trend (strongly suggesting a *word-of-mouth* dominant pattern), and even in those countries showing larger deviations in the growth rate, these fluctuations were not able to substantially perturb the baseline constant trend. Said otherwise, we preferred to interpret these deviations as a consequence of the large volatility (or other undetectable phenomena such as the presence of heterogeneity) which typically characterize early diffusion growth rates rather than the consequence of well-established perturbations. Not including further shocks in the initial phase would then hardly bias the estimate of the imitation rate. Pairwise, we did not include shocks occurring in the last two years of the data window given that all parametric forms considered (F1,F2,F3) always include three parameters.

| Country | Selected Model | q & time to 99th percentile of minimum target | Year of achievement of minimum target | Peak | First shock | | | Second shock | | | Third shock | | |
|---|---|---|---|---|---|---|---|---|---|---|---|---|---|
| | | | | | c1 | a1 | A1 or b1 | c2 | a2 | A2 | c3 | a3 | A3 |
| AUS | F3+F3 | 0.16 (2065) | 2043 | Yes | 0.97 | 16.41 | 17.99 | 0.97 | 21.33 | 2.44 | | | |
| AUT | F3+F3 | 0.26 (2040) | 2032 | Yes | 0.63 | 16.07 | 2.19 | 10.45 | 20.00 | 272.05 | | | |
| BEL | F1+F3+F3 | 0.47 (2032) | 2026 | Yes | 2.89 | 4.66 | 8.09 | 4.22 | 9.17 | 21.37 | 0.47 | 10.71 | -1.12 |
| CAN | F2+F3 | 0.19 (2062) | 2041 | Yes | 0.49 | 16.09 | 9.74 | 8.67 | 21.75 | 161.99 | | | |
| CHE | F3 | 0.15 (2065) | 2041 | Yes | 0.48 | 16.59 | 5.47 | | | | | | |
| CHN | F3+F3 | 0.28 (2047) | 2020 | No | 0.75 | 9.95 | 7.59 | 0.41 | 15.61 | 3.17 | | | |
| DEU | F2+F2+F3 | 0.40 (2030) | 2025 | Yes | 3.20 | 11.23 | 8.20 | 0.30 | 16.43 | 0.95 | 0.47 | 19.78 | -1.07 |
| DNK | F3+F3 | 0.08 (2174) | 2074 | Yes | 3.00 | 17.42 | 840.29 | 3.32 | 21.06 | 52.73 | | | |
| ESP | F3+F3 | 0.16 (2066) | 2026 | Yes | 7.99 | 14.76 | 2122.18 | 1.04 | 17.38 | 7.04 | | | |
| FIN | F1+F3 | 0.13 (2076) | 2070 | No | -1.00 | 19.00 | 22.90 | 5.25 | 22.80 | 251.45 | | | |
| FRA | F3+F3+F3 | 0.16 (2080) | 2048 | Yes | 0.85 | 13.00 | 14.18 | 2.18 | 17.25 | 55.34 | 12.02 | 21.82 | 105.81 |
| GBR | F3+F3 | 0.31 (2044) | 2029 | Yes | 1.75 | 17.28 | 33.06 | 2.43 | 21.43 | 23.18 | | | |
| IND | F3+F3 | 0.25 (2063) | 2040 | No | 1.25 | 6.90 | 27.59 | 0.89 | 12.08 | 3.90 | | | |
| ISR | F2+F2 | 0.13 (2105) | 2056 | Yes | 1.06 | 16.39 | 42.46 | 0.63 | 20.00 | 5.43 | | | |
| ITA | F3+F2 | 0.06 (2190) | 2066 | Yes | 0.97 | 13.80 | 70.93 | 1.11 | 17.78 | 53.40 | | | |
| JAP | F1+F3 | 0.23 (2046) | 2023 | Yes | 1.04 | 3.58 | 8.76 | 0.41 | 19.06 | 3.18 | | | |
| KOR | F3+F3 | 0.16 (2082) | 2052 | Yes | 1.51 | 13.61 | 48.63 | 0.78 | 19.46 | 5.23 | | | |
| MEX | F2+F2 | 0.10 (2113) | 2076 | No | 0.38 | 20.15 | 7.79 | 0.54 | 23.52 | 8.57 | | | |
| MYS | F3 | 0.31 (2037) | 2032 | No | 2.96 | 5.92 | 42.99 | | | | | | |
| NLD | F3+F3 | 0.33 (2035) | 2031 | No | 0.43 | 10.34 | -1.18 | 0.90 | 18.07 | 4.82 | | | |
| NOR | F3+F3 | 0.05 (2182) | 2019 | No | 1.29 | 21.00 | 12.29 | 0.08 | 22.96 | 18.91 | | | |
| PRT | F2+F3+F3 | 0.25 (2044) | 2029 | Yes | 1.48 | 14.78 | 15.05 | 11.42 | 18.93 | 153.95 | 1.90 | 20.66 | 10.89 |
| SWE | F3 | 0.15 (2072) | 2050 | No | 0.39 | 18.09 | 3.44 | | | | | | |
| THA | F3+F3+F3 | 0.11 (2117) | 2017 | No | 2.08 | 7.22 | 163.03 | 4.16 | 10.51 | 181.61 | 0.09 | 13.32 | 29.60 |
| TUR | F3 | 0.29 (2048) | 2031 | No | 0.72 | 14.83 | 8.82 | | | | | | |
| USA | F3+F2 | 0.19 (2053) | 2020 | No | 0.26 | 15.41 | 1.58 | 0.06 | 23.32 | 4.24 | | | |

*Table 1* The GBM fit in the 26 countries considered under the minimum target: summary information. Country (first column), selected model (second column), imitation rate and the year to reach the 99$^{th}$ percentile of the short-term target in the absence of incentives (column 3), year of achievement of the short-term target in the presence of incentives (column 4), presence of peak in annual adoptions (column 5), estimates of the shock parameters (columns 6-8 first shock; 9-11 second shock; 21-14 third shock)

*3.1 Fit to country-level data: the short-term ("minimum") scenario*

We report the results of the "minimum" scenario in greater detail because essentially all countries considered had set a clear short-term policy target at the national level (see the online appendix Table A. 1). Table 1 summarises the selected models and reports the key parameter estimates. Additionally, we reported: (i) the best fits to annual adoptions in all countries considered (Figure 1), (ii) the best fit to annual data with the resulting best forecast up to 2030 (Figure 2), (iii) the fit to the (log-transformed, %) growth rates of cumulative adoptions (Figure 3). The legend in each graph specifies the type of shock functions (in a temporal order) that were selected by model fit. For example, in the case of Australia in Figure 1, "F3+F3" means that both the first and the second selected shocks, visually initiated around 2007 and 2012, belonged to form F3. Full numerical details including confidence limits and goodness-of-fit measures are reported in the online appendix. For ease of exposition we split the results of this section into a number of distinct sub-sections.

### 3.1.1 Fit by the basic Bass model and the innovation coefficient

In essentially all countries the basic Bass model resulted inadequate to reproduce the temporal trends of SPP adoptions, as was clear both graphically and from the values of the multiple partial correlation coefficient $\tilde{R}^2_{i-1,i}$. Nonetheless, the fit by the Bass model revealed the negligible role of the external component in all countries ($\alpha/q < 10^{-4}$), a fact previously noted in Guidolin and Mortarino (2010). This was confirmed by subsequent GBM fits, implying that a generalised internal model (GIM) was fully adequate for the data considered. In substantive terms, this means that the SPP market was launched without a pre-existing steady support from media or public communication i.e., the diffusion effort was entirely sustained by word-of-mouth only.

### 3.1.2 GIM fit: structure of shocks in the different countries

Use of a generalised internal model allowed to satisfactorily reproduce the temporal trends of both annual adoption data (Figure 1) as well as of annual growth rates in all countries (Figure *3*). As for the number of shocks terms, in four countries (Switzerland, Sweden, Turkey, and Malaysia) a single shock was sufficient to adequate fit the data, in other five (Belgium, France, Germany, Portugal, and Thailand) three shocks were necessary, while in the remaining countries two shock functions were sufficient. The selected shock terms resulted essentially always positive and mostly belonging to forms F2 or F3, with the exception of Belgium and Japan, where an F1-shock was detected. No positive increasing shocks were identified during 2006-2016 - they were detected in some countries in Guidolin and Mortarino (2010) for the pre-2006 epoch - as intuitively confirmed by visual data inspection. For Belgium and Germany evidence of a negative shock initiated in the last epoch in the data (Figure 2) was found. Though short lasting, these negative shocks caused the predicted growth rate of adoptions to temporary fall below the level which was expected to occur at the given stage of the lifecycle (given the level of the market potential).

### 3.1.3 GIM fit: predicted profiles of adoption in the different countries

The main fitted shocks were associated, in most countries, to large adoption waves initiated with a surprising synchrony around 2007 (Figure 1 and Figure *3*), irrespective of the scale achieved by the market at that time. These waves are especially apparent from the dramatic post-2007 increase in growth rates (Figure *3*), which was in sharp contrast with the pre-2007 pattern, characterised instead by a fairly long epoch characterised by oscillations around an essentially constant trend (Figure *3*), as is typical of a pure word-of-mouth market in the initial phase of the lifecycle. Exceptions are (Figure *3*) Germany and Japan on the one hand, which were pioneers in setting up robust SPP incentive programs well before the 2000, later followed by Italy, Spain and Korea, and the two "delayers" Turkey and Mexico, where the major shocks occurred a few years later, around 2010.

The factors underlying this synchronous take-off are possibly multiple. They certainly include the documented expansion in public support to SPP (see Table A. 5 in the online Appendix). However, in several cases these measures were established lately under the cogent pressure of the 2008 deadline set by the Kyoto protocol targets about the abatement of greenhouse gases emissions (Kyoto Protocol, 1998), therefore as short-term actions aimed to fulfil standing international commitments in the absence of a well-established national long-term plan. It should also be mentioned the dramatic blow-up of the oil price (a 5-fold increase during 2000-2007, ITRVP 2018) that possibly forced a number of further countries to invest in the SPP technology by imitating the few innovators such as e.g., Germany. Specific situations were also affected by local circumstances, such as the dramatic increase in Japan SPP adoptions in 2013, the year following Fukushima disaster.

Some countries (Australia, Belgium, Denmark, Israel, Korea, Portugal, Spain and Thailand) clearly showed multiple waves in annual adoptions after 2007 (Figure 1). The underlying factors included (Table A. 5 in the online appendix) (i) a lack of coordination in incentive programs between different districts of the same country, as happened in Australia, (ii) a presence of massive sudden adoptions by public utilities, as for Denmark, where the secondary peak is mostly due to the installation of a single public solar park, (iii) public communication announcing a future reduction in the incentive benefit, causing a "run-to-adopt", as documented for France, (iv) lack of coordination and discontinuity in the incentive communication causing - even in the short term - a temporary lack of ability to sustain the adoption flow, as has been the case for Italy, where the availability of incentive funding, though renewed yearly, was always surrounded by large uncertainty (Palmer et al., 2015), and possibly also for the Netherlands (Vasseur and Kemp, 2011).

As for the more recent years, patterns are more articulated. Focusing in particular on the early adopting countries included in Guidolin and Mortarino (2010), there is a clear decline in the fuelling role of incentive. This is apparent not only in absolute terms (Figure 1) but also in relative ones (Figure 3). In particular, only the US were able to keep a growth rate persistently larger than in the pre-2007 period (Figure 3), as mirrored in an annual adoption curve still fast increasing at 2016 (however, it should be pinpointed that the US market started relatively late, with a small adoption scale still at 2010). In a number of countries, namely Australia, Austria, Canada, France, Japan and the UK, annual adoptions showed a sharp decline and growth rates have returned by 2016 to their pre-2007 levels with some evidence of a declining trend. For Germany, we already mentioned the evidence of a negative shock in the last phase. Finally, in Spain and Italy annual adoptions as well as the growth rate of the cumulative adoptions fell to negligible levels indicating a stall in the adoption process. This stall did not require the inclusion of a negative shock because for both these countries the short-term policy target was low enough to allow the post-2007 incentive phase to essentially achieve the target providing at the same time an excellent fit to the data. Notwithstanding the goodness of fit, the dramatic fall of adoptions in both countries is clearly a worrying phenomenon calling for explanations. We conjecture that for Italy, and possibly also for Spain, the market has become "addicted" on subsidies, so that expectations in future incentive programs might now be playing a deleterious role, as argued in the Discussion.

As regards more recent adopters i.e., countries not included in Guidolin and Mortarino (2010), it is important to recall that they either showed lower adoption scales (in some cases negligible, as for Norway and Finland) because still in an initial lifecycle stage, or are currently far more distant from their minimum target (or have set quite low minimum targets). Nonetheless Turkey, Mexico, Sweden and to a lesser extent China, India, Thailand and Korea, are showing by the end of the data window in 2016 growth rates that are still larger than in the pre-2007 period, and no clear evidence of a declining trend, suggesting, overall, a more persistent action of intervention compared to early adopters. On the other hand, Portugal, Denmark and Malaysia showed little persistence and rapid re-alignment to the pre-shock regime.

3.1.4 GIM fit: estimates of the *word-of-mouth* coefficient

In a GIM with shocks occurring after the initial phase, the estimate of the imitation coefficient *q* would be represented by the early growth rate (Figure 3). Our estimates are in good agreement with Guidolin and Mortarino (2010). Substantial inter-country variation in q estimates - up to an order of magnitude – was observed (Figure 4), ranging from a minimum around 5% per year in Norway and Italy up to a maximum of 47% per year in Belgium. A useful related information is represented by the time that would be necessary to saturate the market potential *m* when imitation is the only driving force of adoption (Figure 4, right panel). For example, the time $t_{m,99}$ necessary to achieve the 99 percentile of the minimum target ranged between 31 years for Belgium and 198 years for Italy (Table 1). Given that the true *q* values are likely over-estimated in many countries, because they likely embed the effects of interventions occurred during the early SPP life-cycle, which we deliberately ignored, this result overall suggests that in SPP markets natural communication forces are weak not effective compared to the time scales which are required to respond to the global threats. This provides per se a strong motivation for the need for public incentive to SPP markets.

3.1.5 GIM fit: persistence vs intensity of incentive

The estimates of the characteristic parameters of the shock functions (Table 1) provide valuable summaries about the structure and effects of shocks. Unlike Guidolin and Mortarino (2010), who found a richer combination of shock forms, including negative as well as positive increasing shocks, we found a good degree of homogeneity of the structural characteristic of the shocks occurred during 2006-2016 i.e., essentially all shocks resulted to be positive, mostly obeying forms F2 or F3, and showing low persistence. This allowed us to meaningfully compare the features of the shocks in the countries considered by looking at their key characteristics namely, intensity vs time persistence. To investigate the relationship between these two parameters we refitted the model under a number of homogenising hypotheses: 1) we used only one type of shock forms i.e., F3 only (the only drawback being a slight lack of fit when replacing other types of shocks), taking the estimate of parameter *A* as a measure of *intensity*, and the estimate of ($1/c$) as a measure of *time persistence*; 2) for countries showing well-separate shocks (e.g., Germany, Japan, Korea), documenting distinct incentive actions, we considered estimates from all shocks; 3) for countries showing very close shocks (e.g. the new one arising just one year after the end of the previous one, suggesting an issue of lack of policy coordination rather than genuine different policy actions), we re-fitted a single shock model, in order to provide a feeling of the overall duration of the incentive period; 4) we deliberately disregarded shocks arising in the last two years of the data window. The resulting scatterplot between persistence (horizontal axis) and intensity (vertical axis), showed a marked trade-off (Figure 5). Also notable is the lack of shocks persistence. In fact, in twelve countries the average duration was under one year, in nine countries between one and two years, whereas only in a few cases i.e., Switzerland, Canada, Sweden, Japan, the UK and the US persistence exceeded two years.

3.1.6 GIM fit: achievement of the "minimum" target.

The predicted market growth up to the achievement of the minimum target (Figures 2) –was based on the assumption that no further shocks occur, so that the subsequent market evolution is that of a pure word-of-mouth model. On this assumption, most countries considered would reach the minimum target by the next 10-20 years, in some cases with a further peak. Given the dominant role of the shocks, the subsequent market dynamics is largely constrained by the magnitude of the minimum target relative to the market size at 2016. Note that for Germany and Belgium the further peak is the consequence of the restoration of word-of-mouth after the effects of the negative shock arisen in the last epoch have disappeared, while for other countries as Korea, Malaysia and Mexico, this is instead due the "natural" peak arising in pure word-of-mouth markets. In the case of India, the natural peak is largely overlapped with the last-shock increase. For Italy and Spain – where adoptions suddenly fell to negligible levels after the last incentive phase - the achievement of the minimum target was purely the consequence of the low levels of this target that had been set by the National authorities.

3.1.7 Commonalities in adoption patterns

By focusing on key model parameters, namely the imitation rate $q$ and a summary indicator of the overall shock effects computed as the product eff=$A/c$ of intensity ($A$) times average persistence ($1/c$) resulting from the analyses of section 3.1.5, we attempted at clustering countries together. An application of the K-means algorithm selected as optimal the 4-clusters grouping reported in Figure 6. Notably, while the distribution of the estimated imitation rate is quite dispersed, the distribution of $A/c$ exhibits three outliers, namely Denmark, Italy and Spain taking exceptional values. It so happened that two extreme groups emerge (Figure 6), namely the one including Denmark, Italy, and Spain, characterised by very low $q$ (i.e., a largely ineffective spontaneous agents' communication) and by very intense but short-lasting shocks, and the one including Germany and Belgium, characterised by a very large imitation rate but an average shock efficacy aligned on the average behavior. The remaining countries are partitioned into two larger groups, one with relatively larger imitation rate but a milder shock efficacy, the other one with a lower imitation rate but a larger shock efficacy.

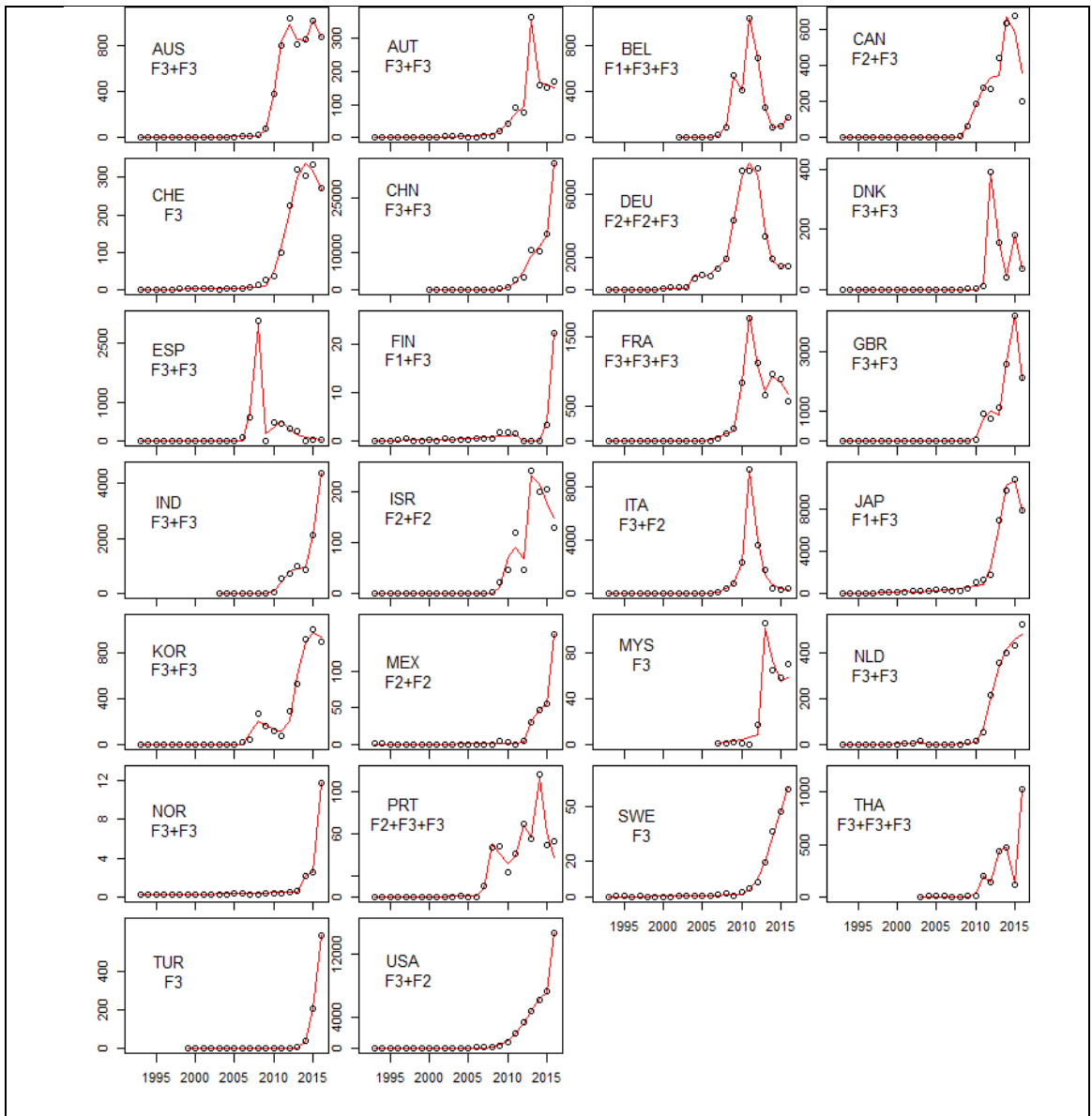

Figure 1. The GBM fit in the 26 countries considered under the minimum scenario: observed vs predicted annual SPP adoptions during 1992-2016. The legend in each graph specifies the type of shock functions identified by model fit e.g., F3+F3 (as in the case of Australia) means that both shocks identified in the data belonged to form F3.

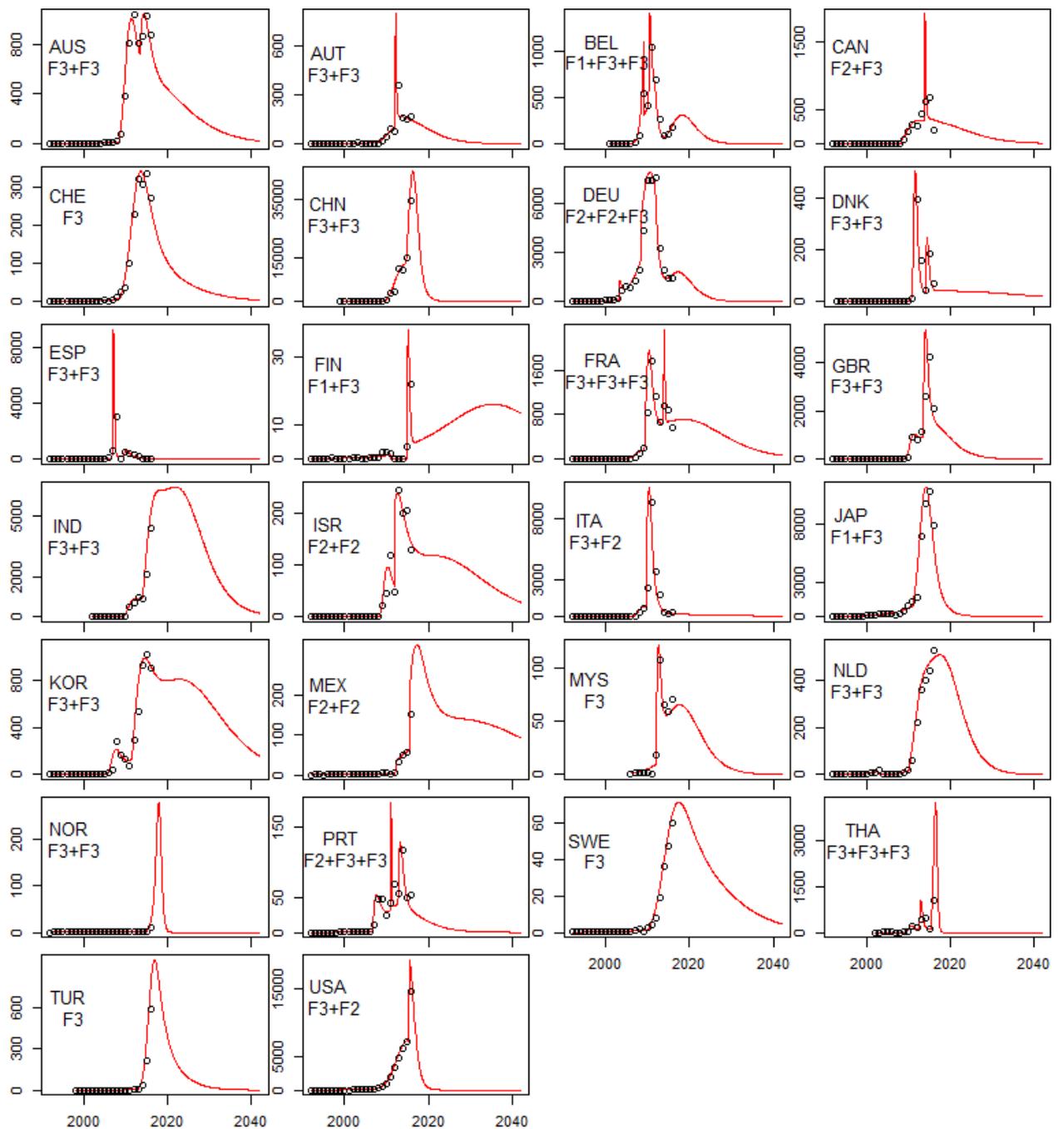

Figure 2. The GBM fit in the 26 countries considered under the minimum scenario: observed vs optimal forecasted annual SPP adoptions curves until 2040. The legend in each graph specifies the type of shock functions identified by model fit. The predicted adoption curve is smoother than in Figure 1 because it has been drawn with a monthly resolution (in Figure 1 we reported observed vs predicted annual figures).

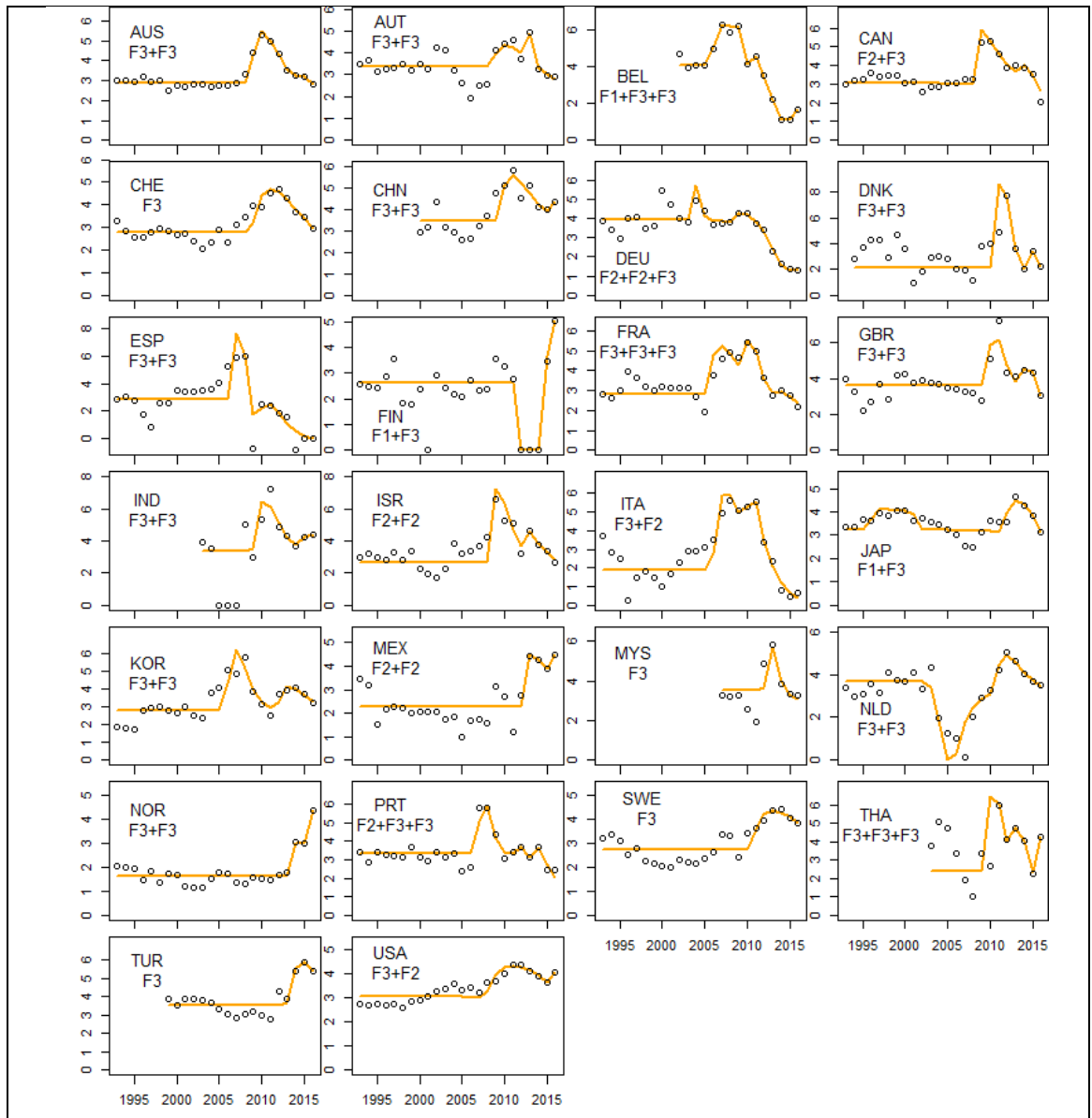

Figure 3 The GBM fit in the 26 countries considered under the minimum scenario: observed vs predicted annual (%) growth rates on adoptions of SPP during 1992-2016 (log scale). The legend in each graph specifies specifies the type of shock functions identified by model fit.

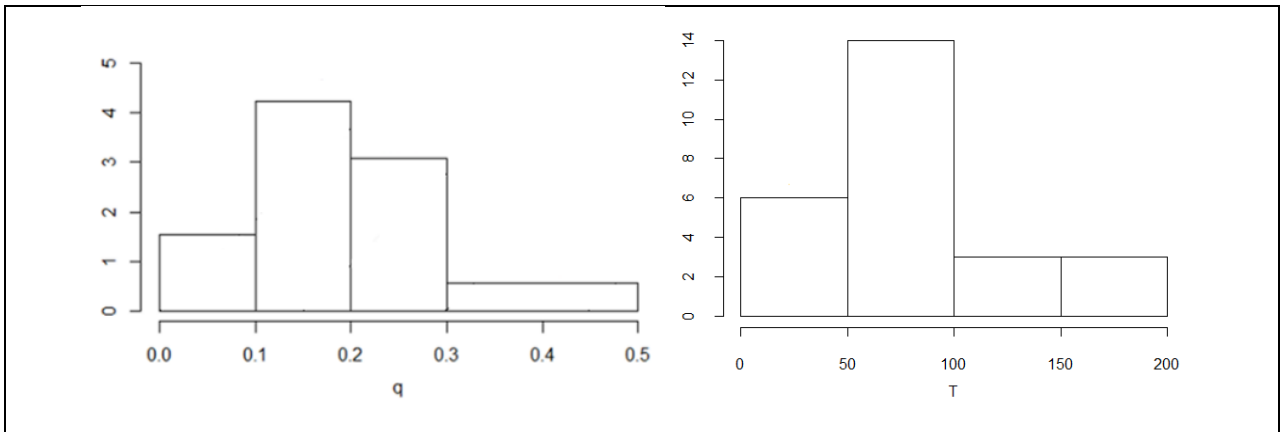

Figure 4. The GBM fit in the 26 countries considered under the minimum scenario. Distribution of the estimates of the imitation rate q and of the time ( $T_{m,99}$) which would be required to reach the 99$^{th}$ percentile of the minimum target in the absence of incentive.

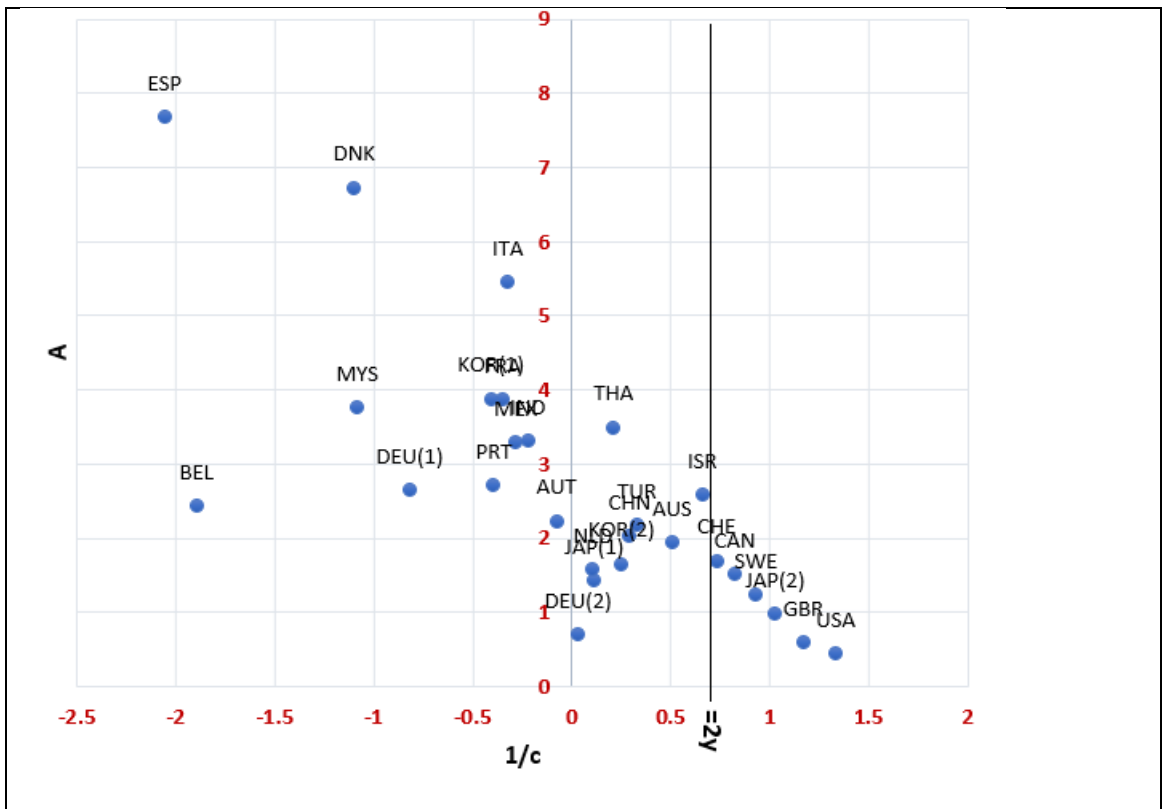

Figure 5. The GBM fit in the 26 countries considered under the minimum scenario. Scatterplot of the best estimates of the parameters (A,1/c) (reported in log-scale) of the F3-form shock functions that best fitted the data. The black line (=2y) represents the threshold of two-years persistence of incentives, whereas the y-axe is the threshold of one-year persistence.

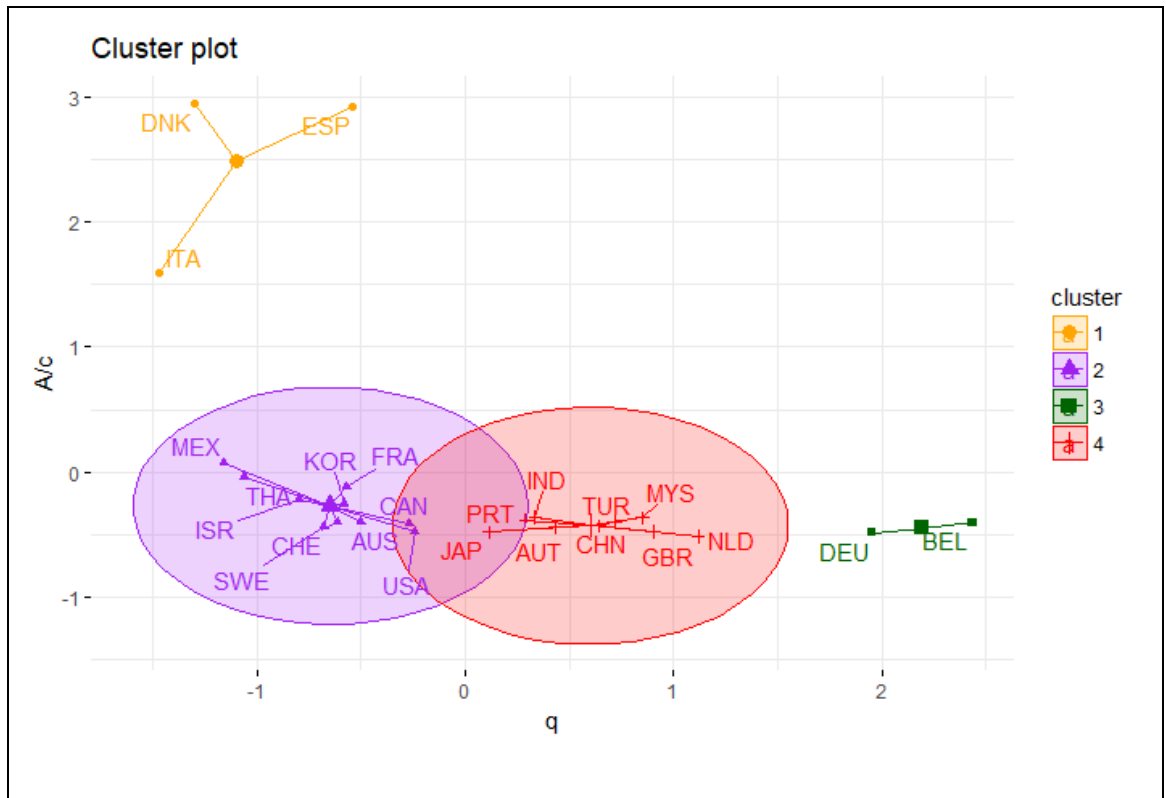

Figure 6. The GBM fit in the 26 countries considered under the minimum scenario. Graphic representation of K-means clustering of countries based on model (standardized) estimates for the imitation rate (q) and for the strength (A/c) of the main shock term resulting from the analyses of section 3.1.5.

*3.2 GIM fit to country-level data: fit of the market potential.*

In internal models, the market potential can be estimated only once the market ends its early exponential phase at speed q and starts bending towards saturation, as clear by rewriting the growth rate equation (6) as follows:

$$m = Y_\infty = \frac{q}{q - r(t)} Y(t)$$

Recalling that $0 < r(t) < q$, previous formula means that as far as market growth is aligned on its early exponential phase at rate q, no prediction of the market potential is possible because the eventual value of the process ($Y_\infty$) is infinity. On the other hand, as $r(t)$ deviates from its early value q, estimation of $m$ gradually becomes feasible though a risk of bias in the estimates of remaining parameters arises (Van den Bulte and Lilien, 1997) unless sufficient data are available.

The previous remark implied that for the majority of countries considered the direct estimation of the market potential was unfeasible because these countries either showed re-alignment to their early growth path at rate q after their last positive shock (as was the case of China, Switzerland, Denmark, Finland, Great Britain, Korea, Israel, Japan, Malaysia, the Netherlands) or were still within an expansive phase (with the growth rate in excess of the level of the early phase) due to the last positive shock, as was the case of India, Mexico, Norway, Sweden, Thailand, Turkey and the US. Even for most remaining countries the evidence of a decline in the growth rate is weak enough - or there are too few data - to reliably estimate the saturation level (see Table A. 3 in the online Appendix). A good example is Canada, where after the post-shock re-alignment to the constant growth path, the very last observation showed a marked drop in both annual adoptions and the growth rate. This would yield an estimate of the market potential of about 3 GW, a value very close to the last observation available (and 50% less than the minimum target), which is clearly trivial due to the large underlying uncertainty. However, this estimate is characterised by a very large confidence interval due to the paucity of data. A similar situation occurred also for Australia. As for the already highlighted situation of Italy and Spain, where both annual adoptions and the relative growth rate dramatically dropped after the disappearance of the last positive shock, the fitted level of *m* is overlapped on the minimum target, which in turn was very close to the last cumulative datum. However, by fixing a level of the market potential only slightly higher than the minimum target allowed to evidence the presence of a strong negative shock in the last phase of the data that needs would need to be reabsorbed in order to achieve market saturation. This further confirms the previous remarks about the anomalous pattern observed in these countries. Eventually, the only case where m could be estimated adequately was Germany, for which, after the disappearance of the negative shock occurred in the last phase a further 30% growth with respect to the level achieved in 2016, is predicted. Details are reported in the appendix.

## 4. Discussion and concluding remarks

Understanding the main determinants and barriers to the diffusion of renewable energies is key for a successful battle against the current global threats, first of all the effects of global climate change. In this article we applied the generalised Bass model (GBM) to the 26 countries that mostly contributed to SPP worldwide adoptions during 1992- 2016, to offer insight on the characteristics of the evolution of the market based on an improved understanding of the role of the main communication drivers of diffusion, namely the media and word-of-mouth, vs the incentive actions from the public system. We paid special attention on the major shocks occurred over the decade 2007-2016, during which the installed SPP capacity worldwide experienced an unprecedented growth, possibly resulting from an epoch of major policy effort aimed to sustain the domestic SPP demand.

Our principal findings were as follows. First, in all countries considered the effects of media communication proved to be negligible, suggesting that the SPP technology started its lifecycle lacking an adequate support by the media system. This confirms previous results by Guidolin and Mortarino (2010), and implies that the SPP markets early growth, say during the first 15 years, was sustained by word-of-mouth only. However, the magnitude of word-of-mouth, though quite heterogeneous among countries, resulted generally low i.e., insufficient to ensure the achievement of any target of market development within the time schedules specified by international agreements. Overall, the previous two findings indicate that the communication forces acting on SSP markets are generally weak, thereby calling for the need for external interventions. Further, most market growth has occurred everywhere in the form of massive positive shocks initiated synchronously 2007, possibly following incentive measures in the various countries. However, the parameter estimates describing the temporal pattern of these main shocks documented a generalised lack of temporal persistence of the effects of incentive, as well as a sharp trade-off between intensity and persistence of the actions namely, the more intense the less lasting they were. The lack of persistence seems to be especially true for the countries that acted as early SPP adopters (say, those included in the analyses by Guidolin and Mortarino 2010), while conditions seems to be slightly better in a few subsequent entrants such as e.g., Turkey, the US, and Mexico. Jointly considered these factors are conferring to the adoption trajectories observed so far a wave-like, "shock-driven", shape where a background steady growth around the imitation rate is from time to time perturbed by external interventions. This particular shape made unfeasible to directly estimate the market potential, in most cases due to the lack of evidence of growth decline. However, in some countries, particularly Italy and Spain, the post- 2007 incentive waves have been more recently followed by a dramatic drop of adoptions to negligible levels. Inferring from this that the market has entered its maturity phase and is approaching its long-term equilibrium, would clearly be a wrong inference caused by the poor predictive ability by Bass-type models in markets dominated by external interventions while still at low scales of adoption, as is the case for the SPP one. Indeed, the achieved scale, as measured by the SPP share of energy consumption, exceeds 10% only for Germany, meaning that the market has large space for further development in all the considered countries. However, the possibility to occupy this space by reaching the larger scale that would be required for successfully confronting with the current global challenges in a reasonable time-horizon obviously depends, among other things, on the future ability to remove the current barriers to adoption, as discussed below.

Our model-based findings yield a number of potentially useful suggestions about barriers to adoptions and future policy interventions. A first one, critical for countries where the adoption of this technology is still in its infancy, will be that of promoting an effective role for media communication. Indeed, in Bass-type models continued media communication is of importance especially in the initial stages of the market to create a large initial cohort of innovators capable to "launch" the word-of-mouth component in an effective manner (Mahajan et al., 2000), a fact so far missing in SPP markets.

A second main point regards the nature of the SPP market and the role of public interventions. The SPP market emerged from our results as a frail and complicate one, where public incentive allowed the market take-off but failed in going beyond their direct short-term effect by providing a sustained momentum to the market in the medium and long-term. As documented, the characteristic temporal trend of the market, dominated by incentive-forced waves followed, in many countries, by low, post-incentive adoptions, suggests that the use of incentive was often badly designed i.e., it produced fast short-term results to respond to external constraints (e.g., the large coordinated adoption wave in 2007 to fulfil the deadlines of Kyoto protocol targets) while lacking a structured longer term perspective, which sounds to be a serious shortcoming.

A further, subtle, barrier might emerge from the discontinuous policies adopted so far, with incentives appearing in a wave-like form. This would make optimal for individuals not to adopt in an incentive-free period while waiting for the next incentive round, and therefore forcing - thanks to their non-adoption behaviour - the Government to provide it. This deleterious role of expectations, which we have analysed in a game-theoretic setup elsewhere (Della Posta et al., 2019), is clearly at risk of being amplified from governments' policies principally aiming at short-term results because missing a long-term perspective. Breaking down this vicious cycle requires restoration of credibility of energy policies in the different countries and possibly, a "forward-guidance" of expectations (Della Posta et al., 2019). We believe that the case of Italy and especially that of Spain are illustrative of this issue of bad design. Indeed, the implementation in

Spain of a very high FIT rate in 2007, resulted in a dramatic boom in adoptions which found the Spanish government unprepared yielding a sudden drop in the subsidies, further reinforced by the advent of the global financial crisis. These circumstances caused the sudden collapse of the Spanish SPP market (Dusonchet and Telaretti, 2010; Movilla et al., 2013). An attempt to remedy the stall by a significant revision of the FIT in 2010, was able to produce only a modest result, possibly confirming that the aforementioned barriers were definitely established.

Notably, the cases of Italy and Spain, on the one hand, and Germany (and Belgium) on the other hand, were detected by a cluster analysis showing that Spain and Italy identified a group by their own characterised by low levels of word-of-mouth and by a parallel resort to very strong (but short-lasting) incentive measures, while Germany and Belgium formed another group on the opposite extremum. This seems to suggest a trade-off in adoption behaviour upon which in countries with weak endogenous communication potential (other things being equal) an excessive use of subsidies was able to bring this low potential to collapse, rather than stimulating it, whereas in countries endowed with a sufficiently robust communication potential the lower need for large shocks prevented such a disruptive impact and, in some cases (e.g., the US), has persistently increased the market growth rate, perhaps also reinforcing endogenous communication. Overall, these results reinforce the idea that SPP markets require public support but also that this support needs being appropriately designed and forward guided.

This work has a number of limitations, lying in both the data and the modelling setup. - Given the importance of the topic, the availability of public data is the most disappointing problem. We had to rely on aggregate data on total installed power because, with a few exceptions, no publicly available harmonized international data on number of installations disaggregated by type of agent (households, vs firms, vs public enterprises and utilities) are currently available. This makes the adopted diffusion models for installed capacity somewhat crude as they postulate homogeneous adopters while these are instead markedly heterogeneous. This heterogeneity is not only intrinsic to the type of agent but also extrinsic i.e., related to the nature of the "durable" goods that are the object of the adoption. A first level where this problem becomes apparent is the lifespan of the durable, which ranges from 15-25 years for a household, up to a much longer period for public utilities for whom switching towards RE represents, then, a very-long-term investment. At this level one might remark that SPP adoption by public utilities would anyhow deal – simply on a very large scale – with the same technology available to households, so that a use of the Bass model in the short-medium term shouldn't substantially violate its basic hypotheses (such as being a "first-purchase" model). However, when referring to time horizons longer than those considered here, things will obviously be different, especially if rapid technological progress will lead to new generations of the technology, requiring to consider models for repeated adoptions, along the strain originated by the Norton-Bass model (Norton and Bass, 1987). Still in relation with the data issue, clearly the differential equation model which is applicable to atomistic individual' decisions, is hardly correct for public utilities, whose adoption model is rather represented by a marked stochastic point process. Further, the generalised Bass model keeps a constant market potential $m$ as in the basic Bass model. Clearly, the process model for the market size in real SPP markets is far more complicate due to its long time scales. These time scales imply both demographic effects in the population of potential adopters as well as effects following from the increasing competitive pressure on the price of the technology, which softens budget constraints and, other things being equal, should eventually expand the population of potential adopters. Therefore, the inclusion of a temporal trend (say, m(t)) for the market potential, be this exogenous (see e.g., Centrone et al., 2007) or endogenous, might be more appropriate. Here we opted for Occam's razor parsimony by setting *m* as a known constant (determined by SPP policy targets in the different countries) simply because both of the previous options would have required including further unknown parameters in the absence of a clear understanding of the underlying process. Moreover, the present models suffer the limitation of following a single technology, thereby disregarding the competition with other options, suggesting that a framework for the competition between different technologies should instead be considered (Guidolin and Tapan 2019).

Any of the aforementioned limitations represent areas of future research. In addition to these general areas, there are the many specific questions that have emerged by the results of this work. A first one lies in the nature of the early evolution of SPP markets, primarily driven by word-of-mouth at an essentially constant growth rate, notwithstanding the marked decline in the price of the technology. We conjecture that this might have been the consequence of the initial presence of more active individuals (perhaps with a stronger environmental sensibility i.e., a heterogeneity effect) who adopted first and were subsequently replaced by new, less active, cohorts, which were still prone to invest but required easier financial conditions under the aforementioned role of expectations. A second, quite related, issue is about the determinants of the magnitude and heterogeneity of imitation rates i.e., word-of-mouth, as the key baseline trigger of SPP markets. Whether this might depend more on socio- economic and cultural factors in the different communities considered (e.g., a more developed environmental sensibility, or a deeper social capital endowment in a higher income setting) or rather reflect different levels of state intervention which mainly affected the size of the imitation rate would require different types of analyses. The former conjecture would obviously yield negative future perspectives for countries where the underlying settings are sub-optimal. A further one deals with the causes underlying the full failure of media communication in supporting SPP markets. An investigation of this issue clearly needs better data to deep to what extent continued media support might be reflected (though hidden) in the values of the imitation coefficient.

To sum up, the SPP market is key for contrasting the current global threats but its frailty is a worrying fact whose determinants need careful investigation and appropriate policy actions. This research has examined the frailties on the communication side and can provide some useful additional information to e.g., statistical and econometric studies focusing on the determinants of SPP adoption by regression models.